\documentclass[prd,twocolumn,  amsfonts, amsmath, bbm, amssymb, superscriptaddress, showkeys]{revtex4-2}

\usepackage{graphicx}

\usepackage[normalem]{ulem}
\usepackage{physics}
\usepackage{tikz,lipsum,lmodern}
\usepackage[export]{adjustbox}
\usepackage{soul}

\usepackage[unicode=true,pdfusetitle,
 bookmarks=true,bookmarksnumbered=false,bookmarksopen=false,
 breaklinks=false,pdfborder={0 0 0},backref=false,colorlinks=true,citecolor=blue,urlcolor=violet 
]{hyperref}
\usepackage{xcolor}
\newcommand{\orcid}[1]{\href{https://orcid.org/#1}{\includegraphics[width=10pt]{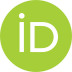}}}

\newcommand{\bket}[1]{\mbox{$|{#1}\rangle$}}

\newcommand{\qup}{\mbox{$\bket{\uparrow}$}}
\newcommand{\qdwn}{\mbox{$\bket{\downarrow}$}}

\begin{document}
\title[]{Emergence of Classicality in Stern-Gerlach Experiment via Self-Gravity}
\author{Sourav Kesharee Sahoo\orcid{0000-0003-1812-0417}}
\email{sourav.sahoo1490@gmail.com.} 
\affiliation{Department of Physics, BITS-Pilani K K Birla Goa Campus, Goa-403726, India.}
\author{Radhika Vathsan\orcid{0000-0001-5892-9275}}
\email{radhika@goa.bits-pilani.ac.in} 
\affiliation{Department of Physics, BITS-Pilani K K Birla Goa Campus, Goa-403726, India.}

\author{Tabish Qureshi\orcid{0000-0002-8452-1078}}
\email{tabish@ctp-jamia.res.in}
\affiliation{Center for Theoretical Physics, Jamia Millia Islamia, New Delhi 110025.}

\begin{abstract}
Emergence of classicality from quantum mechanics, a hotly debated topic, has had no  satisfactory resolution so far. Various approaches including decoherence and gravitational interactions have been suggested. In the present work, the Schr\"odinger-Newton model is used to study the role of semi-classical self-gravity in the  evolution of massive spin-1/2 particles  in a Stern-Gerlach experiment. For small mass,  evolution of the initial wavepacket in a spin superposition shows a splitting in the magnetic field gradient into two trajectories as in the standard Stern-Gerlach experiment. For larger mass, the deviations from the central path are less than in the standard Stern-Gerlach case, while for high enough mass,  the wavepacket does not split, and instead follows the classical trajectory for a magnetic moment in inhomogeneous magnetic field. This indicates the  emergence of classicality due to self-gravitational interaction when the mass is increased. In contrast, decoherence which is a strong contender for emergence of classicality,  leads to a \emph{mixed state} of two trajectories corresponding to the spin-up and spin-down states, and not the classically expected path. 
The classically expected path of the particle probably cannot be explained even in  the many-worlds interpretation of quantum mechanics. 
Stern-Gerlach experiments in the macroscopic domain are needed to settle this question.
\end{abstract}
\maketitle

\section{Introduction}
Despite its enormous success, quantum theory still has certain unresolved problems, one of which is the emergence of  classicality  \cite{schlosshauer2007decoherence,joos2013decoherence}. Quantum theory characteristically  involves superpositions of states, which are entirely absent in the classical world. If one believes that quantum theory is  fundamental, it should lead to the classical world in some limit. Quantum theory in itself provides no  mechanism for the emergence of classicality. 

\begin{figure}[t]  
\centering
\includegraphics[width=0.845\columnwidth]{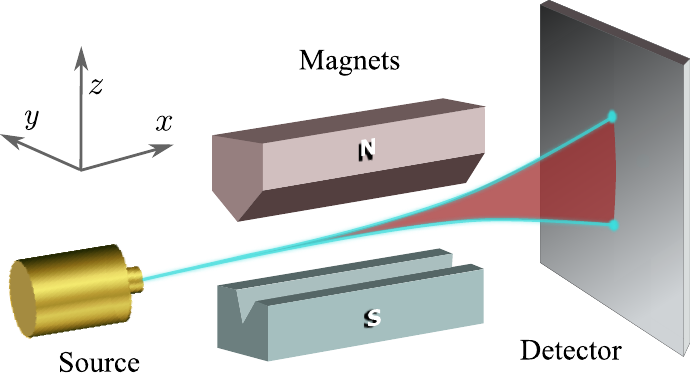}

         \caption{Schematic representation of a typical Stern–Gerlach setup:  quantum  spin-1/2 particles are deflected in one of two directions by an inhomogeneous magnetic field (blue lines) whereas  classical  magnetic moments are deflected by an amount proportional to the $z$-component of their spin (red). }
         \label{sgsetup}
\end{figure}

Consider the  Stern-Gerlach experiment (see Fig.~\ref{sgsetup}),  a prototype example of a quantum measurement \cite{bohm1989quantum}, and one that bears the distinction of demonstrating the fundamentally quantum nature of spin. A spin-1/2 particle  subject to a magnetic field gradient experiences a potential proportional to the component of the spin along the magnetic field direction.  
A general quantum state of a spin-1/2 particle is  a superposition  $\cos\frac{\theta}{2}\qup + \sin\frac{\theta}{2}\qdwn$ in the  spin-$z$ basis.In  a magnetic field gradient, the trajectory of this particle
splits into two, and it always lands in one of two spots on the screen. This general state is also an eigenstate of $\hat{S}_{\theta}=\vec{S}\cdot \hat{n}$, the spin component  along the direction  $\hat{n}$ at an angle $\theta$ to the $z$-axis. Classically this would represent an angular momentum  vector pointing along $\hat{n}$. The force on the classical magnetic moment in a field gradient is proportional to $\cos\theta$, so that the particle  follows a single trajectory landing on the screen  somewhere between the spots corresponding to the spin states $\qup$ ($\theta = 0$) and $\qdwn$ ($\theta=\pi$), with a continuous set of possibilities. 
The question then is  whether  this classical behaviour can emerge from quantum dynamics if the particle is massive enough. To our knowledge a Stern-Gerlach experiment with macroscopic particles  has not yet been performed, and whether it would lead to the classical result remains unverified.

A major contender  among theories for the emergence of classicality is  decoherence\cite{joos2013decoherence,schlosshauer2007decoherence},  which emphasizes the  unavoidable role played by the environment in the evolution of a quantum system. 
Examining the experiment in the context of decoherence, the initial superposition of pure spin states evolves into a mixed state, which is an incoherent superposition of  the two trajectories corresponding to those of $\qup$ and $\qdwn$ \cite{Venugopalan_Kumar_Ghosh_1995}. This predicts that the particle will still land in one of the two spots on the screen corresponding to $\qup$ and $\qdwn$.  Decoherence does not lead to any  intermediate trajectory corresponding to the \emph{classically expected} behaviour of a  particle with magnetic moment passing through a Stern-Gerlach setup.

We wish to emphasize here that in various other experiments, such as  a massive particle passing through a double-slit, decoherence leads to suppression of  superpositions  of the two alternate paths, and  classically expected behavior emerges. In the Stern Gerlach experiment, from classical dynamics one would expect a single outcome corresponding to the $z$-component of the magnetic moment.
However, such an outcome is not explained by decoherence or by the collapse postulate, both of which lead to the emergence of one of the  spin eigenvalues with probabilities predicted by the Born rule.

Following Roger Penrose's proposal \cite{penrose1996gravity}
that gravity could play a role in the emergence of classicality in
massive particles, several authors have investigated the effect of gravity on quantum systems in various ways \cite{colella1975observation,grossardt2016effects,grossardt2016approximations,singh2015possible,yang2013macroscopic}. Here we explore the  role of self-gravity of a massive particle in the emergence of classicality in a Stern-Gerlach experiment, using the  Schr\"{o}dinger-Newton (S-N) equation, earlier introduced by Di\'{o}si \cite{DIOSI1984199}.

\section{The Stern-Gerlach experiment and self-gravity}
\subsection{The Schr\"odinger-Newton equation}
The seeds of the Schr\"odinger-Newton equation came from semi-classical gravity independently considered by M\"oller \cite{moller1962theories} and Rosenfeld \cite{rosenfeld1963quantization}. In this approach, quantized matter is assumed to be coupled to the classical gravitational field \cite{mattingly2005quantum,kibble1981semi,bahrami2014schrodinger}, and the expectation value of the energy-momentum tensor with respect to the quantum state $|\Psi\rangle$ of matter is the source in the  Einstein field equations:
\begin{equation}
G_{\mu\nu} = \frac{8 \pi G}{c^4} \langle \Psi | \hat{T}_{\mu \nu}| \Psi \rangle.
\label{e2}
          \end{equation}
In the Newtonian approximation,  this naturally leads to the Schr\"odinger-Newton equation \cite{bahrami2014schrodinger,van2011schrodinger,salzman2005investigation,giulini2012schrodinger},
 \begin{equation}
          \Bigg( \frac{p^2}{2m} + V_{G}\Bigg) \Psi(r,t) = i \hbar \frac{\partial \Psi (r,t)}{\partial t}, \label{eq:SN}
      \end{equation}
where
  \begin{equation} V_{G} = - G m^2 \int \frac{|\Psi (r',t)|^2}{|r - r'|} d^3 r',
  \label{VG}
  \end{equation}
is the potential due to gravitational self-interaction, derived from the spatial probability distribution of the  massive particle.
Since this potential depends on the wavefunction itself, the S-N equation is  a \emph{non-linear} modification
of the Schr\"odinger equation. The non-linearity breaks the unitarity of  Schr\"odinger evolution, and opens up the potentialities of  effects that were precluded by  the linearity of quantum dynamics,  including perhaps a dynamical reduction of the wavefunction. Issues caused by the sacrosanct quantum-mechanical linearity being broken have been debated in the literature \cite{anastopoulos2014problems}. However, the hope behind this approach is that the dynamics will be linear for all practical purposes  at the scales at which quantum mechanics has been successfully tested, and the non-linearity will show up only when one approaches the classical limit. There have also
been several  other approaches in which non-linearity has been introduced in order to obtain wave-function collapse \cite{bassi2013models,ghirardi1990markov,adler2007collapse}. The S-N equation, however, is elegant because no adjustable parameters are introduced, and it should naturally lead to the scale at which classicality might emerge.

\subsection{Formulation of the Stern-Gerlach problem}
Consider a Stern-Gerlach experiment as shown in Fig.~\ref{sgsetup}, in which a spin-1/2 particle  of mass $m$ travels along the $x$ axis,  experiencing a magnetic field along the  $z$ axis, with a constant gradient $B_0\hat{z}$
\footnote{In reality the magnetic field should satisfy $\frac{\partial B_y}{\partial y} + \frac{\partial B_z}{\partial z} = 0$. So there should be an inhomogeneous component in the $y$ direction too. In addition there should be a homogeneous component in the $z$ direction. Since these are of no consequence in the dynamics of the particle in the Stern-Gerlach experiment, we ignore them in the analysis here.}.
If we assume that the spin-1/2 magnetic moment $\vec{\mu}$ is due to a single unpaired electron in the particle, then the potential experienced by the particle is given by
\begin{equation}
 V_B(z) = -\vec{\mu}\cdot\vec{B}
     = -\mu_B B_0 z \sigma_z,
     \label{eq:VB}
\end{equation}
 where $\mu_B$ is the Bohr magneton, and $\sigma_z$ is the Pauli spin-$z$ operator. This potential causes a spin-dependent deviation along  the $z$-direction. Assuming an initial constant velocity $v\hat{x}$, the  dynamics of the particle along the $x$-axis is trivial,  just  translating the $x$-position of the particle by $x = vt$  in a given time $t$. So we do not consider the motion of the particle along the $x$ axis explicitly, and just focus on its dynamics along the $z$ axis. The quantum dynamics of the particle in the $z$ direction is  given by the one-dimensional Schr\"odinger equation  under the influence of the potential $V_B(z)$ of  Eq.~(\ref{eq:VB}).

Imposing the  additional self-gravitational potential given by Eq.~(\ref{VG}), the effective dynamics in the $z$ direction is governed by
the S-N equation reduced to one dimension\cite{sahoo2022}, along with $V_B(z)$:
\begin{eqnarray}
          \Bigg(   \frac{p_z^2}{2 m} 
          + V_G(z)  + V_B(z) \Bigg) \Psi(z,t)
                    = i \hbar \frac{\partial \Psi (z,t)}{\partial t}, \label{eq:SNB}
      \end{eqnarray}
with $V_G(z)=- G m^2 \int \tfrac{|\Psi (z',t)|^2}{|z - z'|} d z'$.
The state of the particle in general is given by a spinor $ \bket{\Psi}$, which in the position basis, is a two-component wave-function in a superposition of spin-up and spin-down states:
\[\ip{z}{\Psi(t)} =
\chi_+(z,t) \qup +  \chi_-(z,t) \qdwn.\] 
The position-space probability distribution of the particle required to calculate the gravitational potential is given by
\begin{eqnarray}
|\Psi (z,t)|^2 &=& |\ip{ z}{\Psi}|^2 \nonumber \\
   &=& |\chi_+(z,t)|^2 + |\chi_-(z,t)|^2,
\end{eqnarray}
where the cross terms between $\chi_{\pm}(z,t)$  vanish due to the orthogonality of $ \qup$ and $\qdwn$. Using Eq.~(\ref{VG}), the self-gravitational potential becomes
 $V_G(\Psi) =V_G(\chi_+)+V_G(\chi_-)$, where
$
  V_G(\chi_\pm) =- Gm^2\int \frac{|\chi_\pm (z',t)|^2}{|z-z'|} dz'.
$
The S-N equation for our system can now be expressed as
\begin{equation}
\begin{bmatrix}
-\tfrac{\hbar^2}{2m}\partial_z^2+V_G - \gamma z  & 0\\
0 & -\tfrac{\hbar^2}{2m}\partial_z^2+V_G + \gamma z  
\end{bmatrix}\begin{bmatrix}\chi_+ \\ \chi_- \end{bmatrix} = i\hbar \begin{bmatrix} \Dot{\chi}_+ \\ \Dot{\chi}_-
\end{bmatrix}.
\label{SGSN}
\end{equation}
This represents  two equations for the time evolution of the two components $\chi_{\pm}$, that are {\em coupled nonlinear} equations since the potential term $V_G$ depends on both $\chi_{+}$ and  $\chi_-$. Here $\gamma=\mu_B B_0$ represents the magnetic force in our model.
Introducing a length scale $\sigma_r$  in terms of which  time and mass scales are defined as 
\begin{equation}
t_r=\left(\frac{\sigma_r^5}{G \hbar}\right)^{\frac{1}{3}},~~~
m_r=\left(\frac{\hbar^2}{G \, {\sigma_r}}\right)^{\frac{1}{3}},
\label{eq:scalefactors}
\end{equation}
we work in terms of the dimensionless variables
\[
\tilde{z}=z/\sigma_r,~~~
\tilde{m}=m/m_r~,~~~
\tilde{t}=t/t_r~ , \]
and  rescaled wavefunctions
$ \tilde{\psi} = \sqrt{\sigma_r} \psi,~~
\tilde{\chi}_{\pm} = \sqrt{\sigma_r} \chi_{\pm}. $
 The dimensionless  magnetic force is 
 \begin{equation}
  \tilde{\gamma}= \frac{\gamma}{m_r\sigma_r/t_r^2}
   = \frac{\mu_B B_0}{m_r\sigma_r/t_r^2}.
\label{mforce}
 \end{equation}
Eq.~(\ref{SGSN}) can be rewritten in dimensionless terms  as 
\begin{eqnarray}
          - \frac{1}{2\tilde{m}}\frac{\partial^2\tilde{\chi}_\pm}{\partial\tilde{z}^2}  
          \mp \tilde{\gamma}\tilde{z} \tilde{\chi}_\pm 
          - \tilde{m}^2\tilde{\chi}_\pm \int \tfrac{|\tilde{\chi}_+ (\tilde{z}',\tilde{t})|^2 + |\tilde{\chi}_- (\tilde{z}',\tilde{t})|^2}{|\tilde{z}-\tilde{z}'|} d\tilde{z}' \nonumber \\ 
          = i  \frac{\partial \tilde{\chi}_\pm}{\partial \tilde{t}}. 
          ~~~~
          \label{SGSN-d}
      \end{eqnarray} 

The S-N equation for a spin-1/2 particle  in a Stern-Gerlach setup has been studied  analytically by Großardt \cite{Grossardt_2021}, in the context of spin interference. We have carried out comprehensive numerical simulations  of the problem to avoid being constrained by the limits of analytical approximations. We used the Crank-Nicolson method \cite{Crank_1947,smith1985numerical} to solve Eqns~\eqref{SGSN-d}, with  spatial and temporal grid sizes chosen as $\Delta \tilde{z} = 0.05$ and $\Delta \tilde{t} = 0.01$ respectively, which satisfy the CFL (Courant–Friedrichs–Lewy) condition for stability of the solutions: $\frac{\Delta\tilde{t}}{\Delta \tilde{z}} < 1$. To tackle the singularity in the self-gravity potential, we regularize the integral in the potential using a small dimensionless $\delta$:
$-m^2\int \frac{\tilde\chi_\pm (\tilde{z}',\tilde{t})|^2}{\sqrt{|\tilde{z}-\tilde{z}'|^2+\delta^2}} dz'$. $\delta$ was fixed at 0.01. Fixing the spatial boundaries at $\pm 100$ avoided boundary effects. 

The  initial state of the particle was a Gaussian wave-packet of half width $\epsilon$, localized at $\tilde{z}=0$,  with the spin state in a general superposition $\cos\frac{\theta}{2} \qup + \sin\frac{\theta}{2}\qdwn $, so that 
\[\tilde{\chi}_+(\tilde{z},0) = \cos\tfrac{\theta}{2}\frac{ e^{-\tilde{z}^2/2\epsilon^2}}{\epsilon\sqrt{\pi}}, ~~
\tilde{\chi}_-(\tilde{z},0) = \sin\tfrac{\theta}{2}\frac{e^{-\tilde{z}^2/2\epsilon^2}}{\epsilon\sqrt{\pi}}.\]

\section{Results and Analysis}
\subsection{Dynamics of wavepackets}
\begin{figure}
\includegraphics[width=1.0\columnwidth]{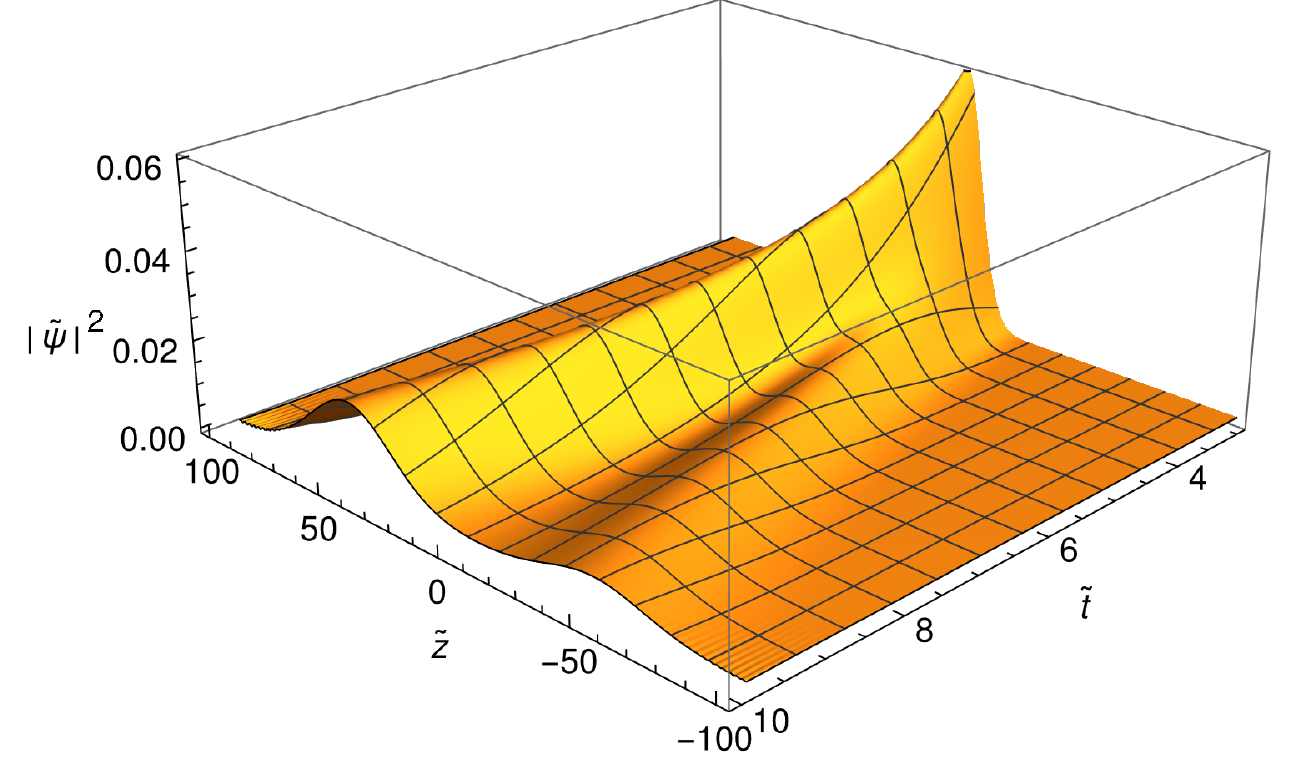}
\caption{Time evolution of $|\tilde{\Psi}|^2$ for $\tilde{m} = 0.1$ and $\theta=\pi/3$, displayed from $\tilde{t}=3$ to $\tilde{t}=10$. The initial wave-packet, localized at $\tilde{z}=0$, evolves into a superposition two wave-packets of unequal heights.}
\label{fig:3Dm0.1}
\end{figure}

\begin{figure}
\includegraphics[width=1.0\columnwidth]{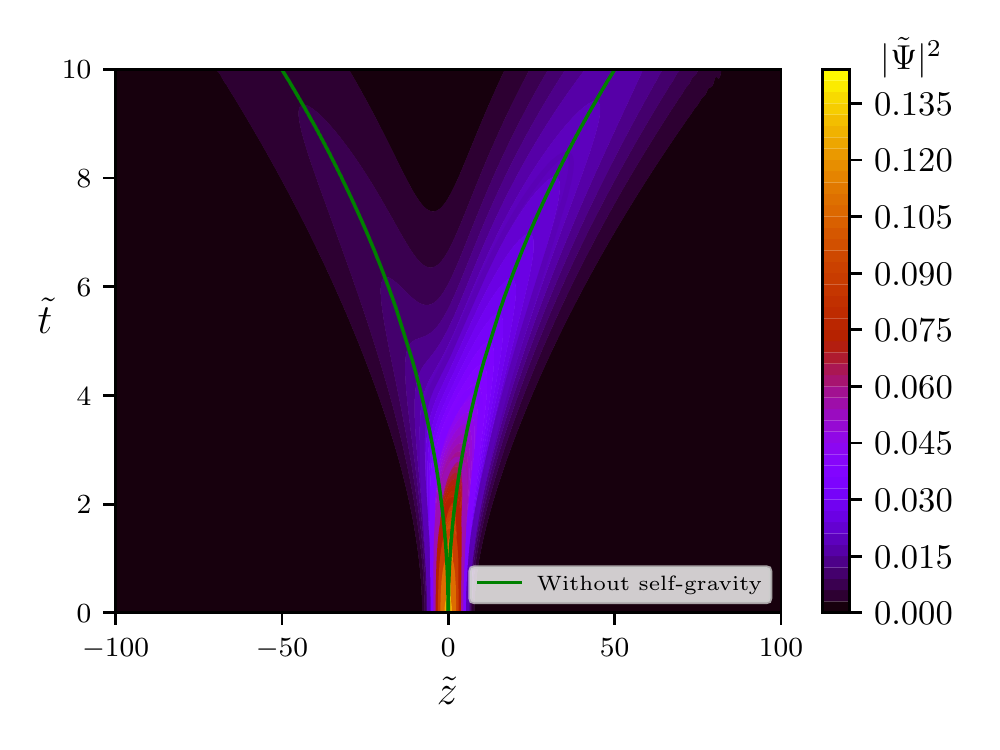}
\caption{Evolution of $|\tilde{\Psi}|^2$ with $\tilde{t}$ for $\tilde{m} = 0.1$ and $\theta=\pi/3$, with self-gravity, represented as a contour plot. The green lines represent the trajectory of the peaks of the two wavepackets without self-gravity. The time evolution in the presence of self-gravity is indistinguishable from pure Schr\"odinger dynamics.}
\label{fig:contourm0.1e4}
\end{figure}

Fig.~\ref{fig:3Dm0.1} shows the time evolution of the probability distribution  for a relatively small mass, $\tilde{m}=0.1$, in an initially asymmetric spin superposition with $\theta=\pi/3$ and $\epsilon=4$.
The initial single spatial wave-packet splits into a superposition of two wave-packets traveling in different directions, consistent with the behaviour expected in the typical Stern-Gerlach scenario. 
A contour plot of 
$|\tilde{\Psi}(\tilde{z},\tilde{t})|^2$ as it evolves in time 
is shown in Fig.~\ref{fig:contourm0.1e4}.
Comparing with the trajectories of the peaks of the spin-up and spin-down wavepackets under pure Schr\"odinger dynamics (shown overlaid in blue),  it is evident that for low enough mass, the presence of self-gravity does little to affect the  pure quantum dynamics,  and quantum superpositions remain unaffected.

\begin{figure}[t]
 \includegraphics[width=1.0\columnwidth]{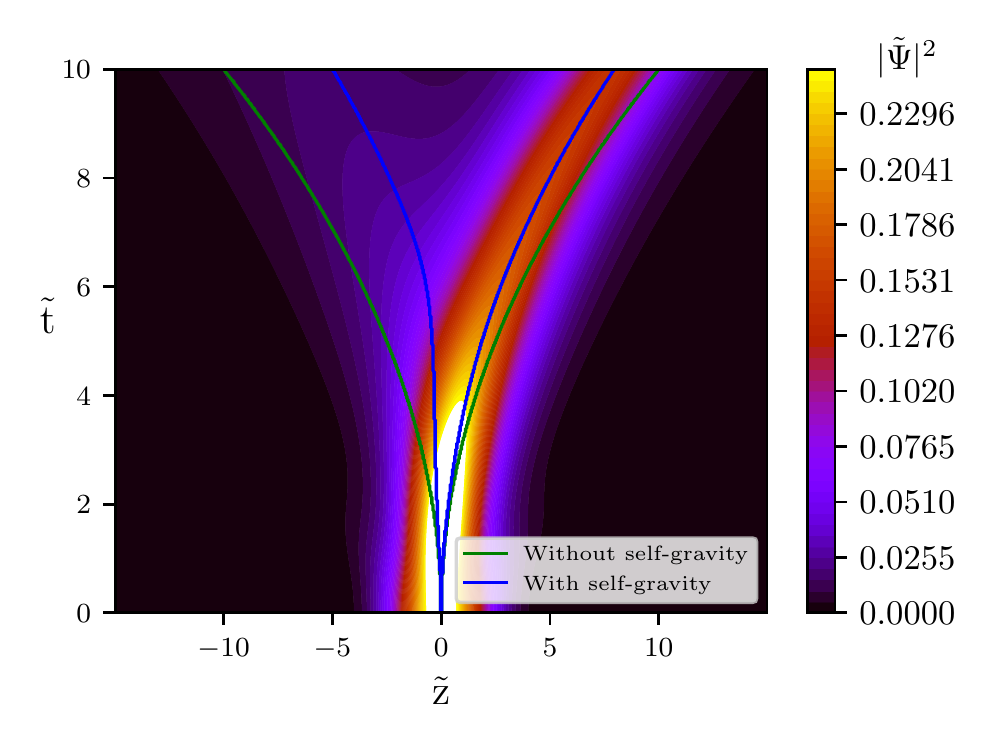}
                \caption{Evolution of $|\tilde{\Psi}|^2$ with $\tilde{t}$ for $\tilde{m} = 0.5$ and $\theta=\pi/3$. Self-gravity leads to an ``attraction" between the two wave-packets. The peaks of the two wave-packets (shown in blue) deviate from the  trajectories of the wave-packets following pure Schr\"odinger evolution (shown in green).}
      \label{fig:contourm0.5}
     \end{figure}

As we increase the mass, the trajectories of the  peaks of the wave-packets begin to deviate from those corresponding to pure Schr\"odinger dynamics. Fig. \ref{fig:contourm0.5} 
displays the contour plot of the probability density in the presence of self-gravity, together with the peak positions of the two wave-packets without self-gravity, for $\tilde{m}=0.5$, $\epsilon=2$. 
The split between the spin-up and spin-down parts of the wavefunction is  less than that  without gravity, which is expected due to the gravitational attraction between the two components. Interestingly, the deflection of the two trajectories is not symmetric about the $\tilde{z}=0$ line when the spin superposition is unequal.

\subsection{Emergence of Classicality}

\begin{figure}[t]
\includegraphics[width=1.0\columnwidth]{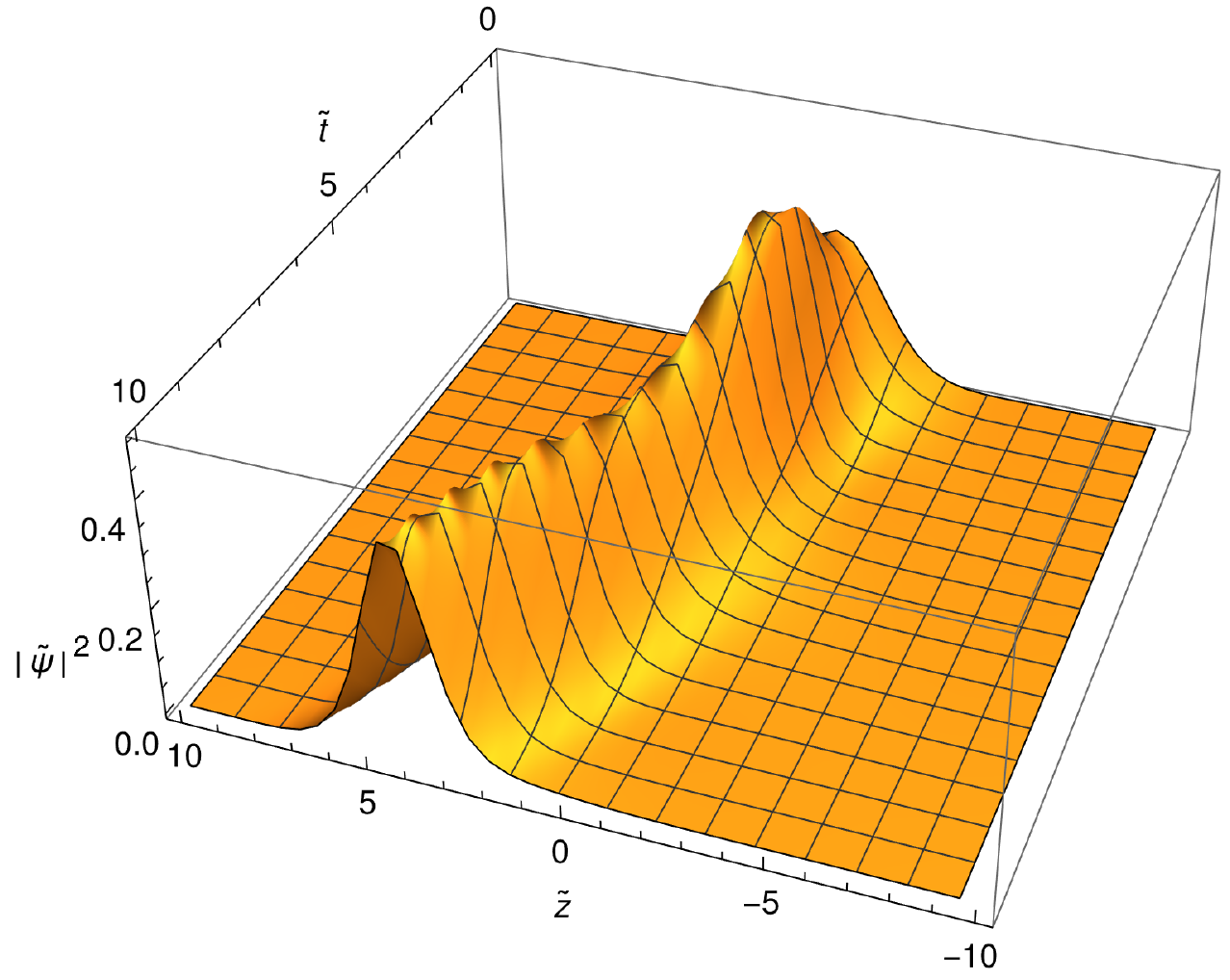}
\caption{Evolution of $|\tilde{\Psi}|^2$ with $\tilde{t}$ for $\tilde{m} = 0.6$ and $\theta=\pi/3$. The initial wave-packet, localized at $\tilde{z}=0$,  continues as a single wave-packet, deviating from the central path due to the magnetic field gradient.}
\label{fig:3Dpsim0.6}
\end{figure}

\begin{figure}[t]
\includegraphics[width=1.0\columnwidth]{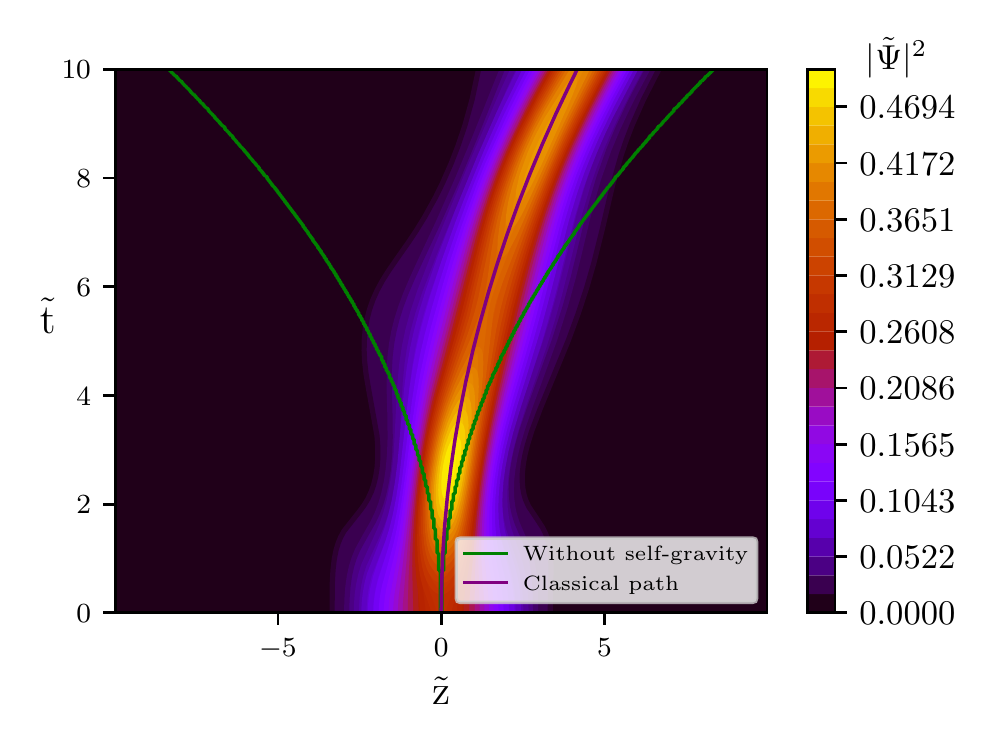}
\caption{Contour plot of the time evolution of $|\tilde{\Psi}|^2$ for $\tilde{m} = 0.6$ and $\theta=\pi/3$. The initial wave-packet, localized at $\tilde{z}=0$, does not evolve into a superposition of two wave-packets. Instead, it continues as a single wave-packet whose center follows the classical trajectory (shown in purple).}
\label{fig:psim0.6}
\end{figure}
The most unexpected results are seen as we increase the mass further. For values of $\tilde{m}$ greater than about $0.6$,  the initial wavepacket does not separate into two, but is deflected along a single path (Figs. \ref{fig:3Dpsim0.6} and \ref{fig:psim0.6}) 
between the two paths of the pure Schr\"odinger case. This implies that in an experiment with an ensemble of massive particles in this spin state, the screen would show a single spot instead of two. 

A classical particle with magnetic moment oriented at an angle $\theta$ with the $z$ axis would have an acceleration $a = ({\tilde{\gamma}}/{\tilde{m}}) \cos\theta$ under the magnetic field gradient, and follow a classical trajectory.
Fig. \ref{fig:psim0.6} shows the peak of the wavepacket for $\tilde{m}=0.6$ closely following this trajectory (shown in purple).
This seems to signal the emergence of classicality, and an apparent absence of quantization of the magnetic moment.  It is  interesting to note that this behaviour appears as a consequence of self gravity,  despite the spin being fully quantized in the analysis. 
  
\begin{figure}[t]
     \centering
          \includegraphics[width=1.0\columnwidth]{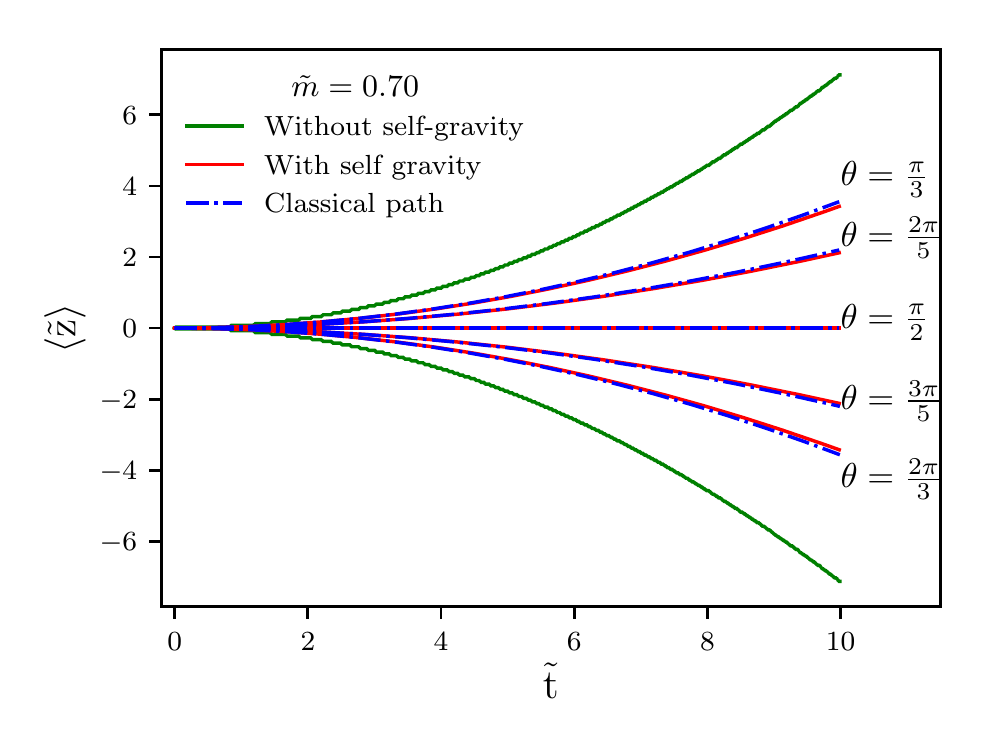}
      \caption{Expectation value of position vs time for  $\tilde{m}=0.7$, for various values of $\theta$. In all cases the trajectory of the peak of the wave-packet follows the classical path reasonably well, indicating the emergence of classicality, driven by self-gravity.}
      \label{fig:peaksm0.7}
     \end{figure}

We tested this effect for various values of  $\theta$, changing the asymmetry in  superposition of the spin states. The results summarized in Fig. \ref{fig:peaksm0.7} show that for high mass ($\tilde{m} = 0.7$), dynamics under the S-N equation recovers the classically expected result in the Stern-Gerlach experiment. Another interesting feature is that the quantum expectation of the position also follows the classical trajectory reasonably well in all the cases studied here. 
The behaviour of the expectation value of the position of the particle indicates that Ehrenfest's theorem \cite{Ehrenfest_1927} continues to hold for the S-N dynamics, despite the nonlinear, self-gravitational potential term. 

\section{Discussion}
Our results indicate that self-gravity could play a role in the emergence of classicality in a spin measurement experiment. As the mass of the particle is increased, the expected deviations of the peaks of the spin-up and spin-down wavepackets are less than for the pure quantum evolution, and for high enough mass the two wavepackets do not separate, but instead travel as one wavepacket along the classically expected path.
This behaviour does not emerge in the decoherence picture.  Neither does it emerge in the many worlds interpretation \cite{mwi1973}. In the latter, the  two wave-packets with different spins, in the superposition state are believed to exist in \emph{two independent worlds}. In each world, the particle is localized in position and has a distinct value of the $z$ component of the spin. The classically expected path of the particle is nowhere in the picture in this interpretation. 
The S-N model thus appears to fare better than other popular candidates for emergence of classicality that do not invoke wave-function collapse. 
However, it should be mentioned that it is not clear if the S-N equation can explain the collapse of a superposition of two distinct states to one of the states, with the probability governed by Born rule. There have been suggestions that the S-N equation may need stochastic modifications to achieve that \cite{Diosi2022}. It has also been pointed out that in the case of two spin-entangled particles, and one of them
passing through a Stern-Gerlach setup, the S-N equation leads to a possibility, at least in principle, of faster-than-light signalling \cite{bahrami2014schrodinger}.


The question remains as to  whether nature actually shows the classically expected path in a Stern-Gerlach experiment with massive particles. Magnetic experiments have been carried out with large atomic clusters, and nano-size particles \cite{Cobalt1991,Arndt2022}. However, in such experiments the magnetic moment of the atomic cluster is also large, and the space quantization may not be apparent.  The experiment has to be performed with spin-1/2  particles in the macroscopic domain. Coming up with large particles with small, stable magnetic moments may be challenging as spin relaxation is often observed in such situations \cite{Heer1990}. 

Our results indicate that the effects of self-gravity in the Stern-Gerlach experiment are visible
for $\tilde{m} \sim 0.3$ onward, on evolution upto  $\tilde{t} \sim 10$. Choice of $\sigma_r$ is guided by  Eqn (\ref{eq:scalefactors}) and experimentally feasible masses.  The smaller $\sigma_r$ is, the larger the mass  for which the  effect of self-gravity will be  noticeable over a relatively short time evolution.  Eqn. (\ref{mforce}) gives the required magnetic field gradient. For example, choosing $\sigma_r=0.371$~nm leads us to $m_r=46.05\times 10^9$~u and $t_r=0.1$~s. This implies that a particle of mass $27.63\times 10^9$~u ($\tilde{m}=0.6$), passing through a  magnetic field gradient of  $28$~mT/m, will display the  emergence classical behavior by time $t=1$~s ($\tilde{t}=10$). These values appear well within the reach of the state-of-the-art technology. Magnetic field gradients up to $80$~mT/m can already be achieved in clinical magnetic resonance imaging (MRI) scanners \cite{Vachha_Huang_2021}.

This experiment will be orders of magnitude less challenging than the Stern-Gerlach-like experiments proposed in the context of gravity induced entanglement, where masses of the order of $6\times 10^{12}$~u and magnetic field gradients of the order of $10^6$~T/m are needed \cite{Bose2017}. Stern-Gerlach experiments for testing the emergence of classicality have an advantage  over interference experiments \cite{sahoo2022} in that  one need not worry  about creating and maintaining coherent superpositions of massive particles. 
The only challenge, we believe, would be in coming up with a particle of such high mass, but having a spin-1/2. Our simulations show that masses lower than about $5\times 10^9$~u will not show any effect of self-gravity in a Stern-Gerlach experiment.
     
\section{Conclusion}
Including semiclassical gravitational self-interaction in the Schr\"odinger equation yields the interesting outcome of emergence of classicality for high masses in the Stern-Gerlach experiment. This is a feature that is amenable to testing in the laboratory, subject to the feasibility of creating spin-1/2 particles of high enough mass.
The question which needs to be settled is whether nature shows the expected classical outcome for large mass particles, or does spin remain quantized all the way to the classical regime. It is not clear if the emergence of a single outcome corresponding to the average spin in a Stern-Gerlach experiment can be explained by any linear theory. Thus this experiment may turn out to be a crucial one in deciding the survival of  nonlinear theories like the S-N equation. 
If the experiments show that for large masses the classically expected result is recovered, it would indicate that there is something lacking in the linear quantum mechanics, including the ideas of decoherence and the many worlds interpretation.

We believe this experimental test should be easier to implement compared to the interference experiments which are being attempted on the mesoscopic scale \cite{arndt2014testing,Fein2019,Bassi_Cacciapuot_2022}.

\begin{acknowledgments}
This work was partially supported by the Department of Science and Technology, India through the grant DST/ICPS/QuST/Theme-3/2019/Q109.
Authors thank Martin Plenio for useful discussions regarding feasibility of the proposed experiment.
SKS would also like to thank  Ashutosh Dash for fruitful discussions and N. K. Patra for the help in 3D plotting. 
\end{acknowledgments}

\medskip
\bibliographystyle{MSP}
\bibliography{References}

\end{document}